\documentstyle[aps,manuscript]{revtex}
\newcommand{\be}{\begin{equation}}
\newcommand{\ee}{\end{equation}}
\begin{document}

\title{ Anisotropy of Ultra High Energy Cosmic Rays in the Dark Matter
Halo Model}

\author{V. Berezinsky}
\address{INFN, Laboratori Nazionali del Gran Sasso, I-67010 Assergi (AQ), Italy
\\ and Institute for Nuclear Research , Moscow, Russia}
\author{A.A.Mikhailov}
\address{Institute of Cosmophysical Research and Aeronomy, 31 Lenin Ave\\
677891 Yakutsk, Russian Federation.}

\maketitle

\begin{abstract}
The harmonic analysis of anisotropy of Ultra High Energy Cosmic Rays is 
performed for the Dark Matter halo model. In this model the relic superheavy 
particles 
comprise part of the Dark Matter and are concentrated in the Galactic halo. 
The Ultra High Energy Cosmic Rays are produced by the decays of these 
particles. Anisotropy is caused by the non-central position of the Sun in the 
Galactic halo. The calculated anisotropy is in reasonable agreement with the 
AGASA data. For more precise test of the model a comparison of fluxes 
in the directions of the Galactic Center and Anticenter is needed.
\end{abstract}

The spectrum of Ultra High Energy Cosmic Rays (UHECR) is measured now 
up to a maximum energy of $(2 - 3)10^{20}~eV$ \cite{HEA,HEFE}. More than 1000  
particles are detected at energies higher than $1\cdot 10^{19}~eV$ 
\cite{UHEHP,UHEA,HEFE,UHEYA}. The detailed energy spectrum was recently 
presented in \cite{UHEAG}. No steepening of the spectrum has been observed 
in the energy range between $10^{18}$ and $2\cdot 10^{20}~eV$. If extragalactic,
the UHE protons must have the Greisen-Zatsepin-Kuzmin (GZK) cutoff 
\cite{GZK} at energy $E_{1/2}=6\cdot 10^{19}~eV$ \cite{BBDGP}. A similar 
cutoff should exist if primaries are extragalactic 
nuclei\cite{BGZ,BBDGP,Roulet} or photons \cite{B70,PB}. 

Recently it was suggested that UHECR can be generated by the decay of 
superheavy  relic particles \cite{KR,BKV,Fram,BiSa}. These particles 
can be effectively produced in the post-inflationary Universe 
\cite{BKV,CKR,KT} and can constitute now a small or large part of 
Cold Dark Matter (DM). As any other form of Cold DM these relic particles are 
concentrated in the halo of our galaxy, and thus UHECR produced by their 
decays do not exhibit the GZK cutoff \cite{BKV}. Realistic 
particle candidates for this scenario and possible mechanisms to provide 
the long lifetime for superheavy particles are discussed in 
Ref.\cite{KR,BKV,Ellis,Yanag}. 

The halo model discussed above has three signatures \cite{BKV,DuTi,BBV}: the 
excess of high energy photons in the primary radiation, direct signal from 
a nearby clump of DM (e.g. Virgo Cluster), and anisotropy  
caused by the asymmetric position of the Sun in the Galactic halo. These 
signatures allow to confirm or to reject the DM halo hypothesis by the data 
of existing arrays.  

As calculations show \cite{BBV,wolf}, anisotropy reveals  
itself very strongly in the direction of the Galactic Center. This prediction 
can be reliably examined by the Pierre Auger detector in the southern 
hemisphere \cite{Cronin}. At present
there is no detector which can observe this direction. In this article we
present the calculations of 
anisotropy for the arrays in the northern hemisphere, taking as an example the 
geographical position of the Yakutsk and AGASA arrays. 

We shall assume that primary photons dominate in the decays
of SH particles as QCD calculations \cite{BK} imply. Then the flux 
of UHE photons in the direction ($\zeta, \phi$) per unit solid angle
can be written as  
\be
 I(\zeta)= K\int_0^{r_{max}(\zeta)}dr \rho_X(R),
\label{eq:flux}
\ee
where $r$ and $R$ are  the distances to a decaying X-particle from the Sun
and the Galactic Center, respectively, $\zeta$ is the angle between the line of 
observation and the direction to the Galactic Center, $\phi$ is the azimuthal 
angle in respect to Galactic plane (the flux depends only on 
$\zeta$), $\rho_X(R)$ is the mass density of superheavy particles ($X$) at 
a distance $R$ from the Galactic Center, $K$ is an overall constant, 
$r_{max}(\zeta)= \sqrt{R_h^2-r_{\odot}^2\sin^2\zeta} +
r_{\odot}\cos\zeta, \;\;\; r_{\odot}= 8.5~kpc$ is a distance between the Sun 
and Galactic Center,   
$R_h$ is the size of the halo, in our calculations
we shall use $R_h=50~kpc$ (the values of 100 and even $500~kpc$ result
in similar anisotropy);
the distance $R$ from the Galactic Center to the decaying particle is 
given by $R^2(\zeta)= r^2+r^2_{\odot} - 2rr_{\odot}\cos\zeta$.

We shall use two distributions of DM in the halo: one given by the
Isothermal Model (ISO) \cite{Prim},
\be
\rho(R)=\frac{\rho_0}{1+(R/R_c)^2},
\label{eq:ISO} 
\ee
and the other -- following the NFW numerical simulation \cite{NFW}
\be
\rho(R)=\frac{\rho_0}{R/R_s(1+R/R_s)^2}.
\label{eq:NFW}
\ee
In the ISO model we shall use for $R_c$ the values $5~kpc$, $10~kpc$ and 
$50~kpc$. For the NFW model the calculations are performed for $R_s$ equal to 
$30~kpc$, $45~kpc$ and $100~kpc$. The NFW distribution \cite{NFW} is 
given in terms of the virial radius $r_{200}$, the rotational velocity  
at the virial distance $v_{200}$, the constant $\delta_c$ and the dimensionless 
Hubble constant $h$. We applied this distribution to our Galaxy using the 
following parameters: the local density of DM 
$\rho_{DM}(r_{\odot})= 0.3~GeV/cm^3$, $v_{200}=200~km/s$ and $h=0.6$. 
As a result we obtain $R_s \approx 45~kpc$. 

The flux (\ref{eq:flux}) was expressed first in terms of galactic coordinates,
longitude $l=\phi$ and latitude $b$, which is given by 
$\cos b= \cos\zeta/\cos\phi$, and then transferred into equatorial 
coordinates, declination $\delta$ and right ascension $\alpha$. 
We calculated the amplitudes of the first and 
the second harmonics ($A_1$ and $A_2$, respectively) and the phase $\alpha$ 
of the first harmonic for the geographical position of the Yakutsk and 
AGASA arrays. These quantities are the standard ones used for measured 
anisotropy.  
The results are given in Table 1 (predictions for the AGASA 
array are shown in brackets). Depending on the parameters of the DM
distribution, the anisotropy varies 
from $10\%$ to $45\%$. The phase of the first harmonic $\alpha \approx 
250^{\circ}$ is close to the RA of Galactic Center, 
$\alpha \approx 265^{\circ}$. 

After this paper was submitted for publication we receved the preprint by 
Medina-Tanco and Watson \cite{Watson}, where similar calculations were 
performed. The results of both calculations are displayed in Fig.1 for 
$E>4\cdot 10^{19}~eV$ together with the data of AGASA (AG) and Yakutsk (YK) 
arrays. The AGASA anisotropy is taken from analysis of Ref.\cite{Watson}.
Our calculations (BM) agree well with that of Ref.\cite{Watson}. Both agree 
with the data of AGASA array  and do not contradict to the Yakutsk data.

{\bf Acknowledgements}\\*[2mm]
The authors are grateful to P.Blasi, M.Hillas, B.Hnatyk, M.Nagano, P.Sokolsky, 
A.Vilenkin, and A.Watson for discussions and correspondence.

\newpage

\begin{table}[t]
\caption[abc]{ Anisotropy}
\center{
\begin{tabular} { c c c c | c c c c }
 \multicolumn{4}{c}{ISO model} & \multicolumn{4}{c}{NFW model}\\
 $R_c$  & $A_1$ & $A_2$  & $\alpha$  & 
     $R_s$  & $A_1$ & $A_2$  & $\alpha$  \\
\hline
5 kpc & 0.43(0.46) & 0.11(0.13) & $250^{\circ}$ & 30 kpc & 0.41(0.45) & 0.10
(0.13) & $250^{\circ}$ \\ 
10 kpc&0.32(0.35)  & 0.06(0.07) & $250^{\circ}$ & 45 kpc & 0.37(0.41) &0.09
(0.11)  & $250^{\circ}$ \\
50 kpc&0.15(0.15)  & 0.01(0.01) & $250^{\circ}$ &100 kpc & 0.33(0.36) &0.07
(0.08)  & $250^{\circ}$\\
\end{tabular}}
\label{tabsig}
\end{table}
\vskip 1cm
\begin{center}
FIGURE CAPTIONS
\end{center}
\vskip 3mm
{\bf Figure 1}: Amplitude and phase of the first harmonic of anisotropy 
for the AGASA and Yakutsk arrays. Solid lines are for the ISO distribution of 
DM and dots (NFW) -- for the NFW numerical simulation. BM and TW show the 
results of this paper and Ref.\cite{Watson}, respectively. The three dots 
of BM (from left to right) are given for $R_s = 30,\;\;45\;\;$ and $100~kpc$,  
respectively; five dots of TW -- for $R_s=10,\;\;20,\;\;30,\;\;50$ and 
$100~kpc$. The AGASA data are taken from Ref.\cite{Watson}.

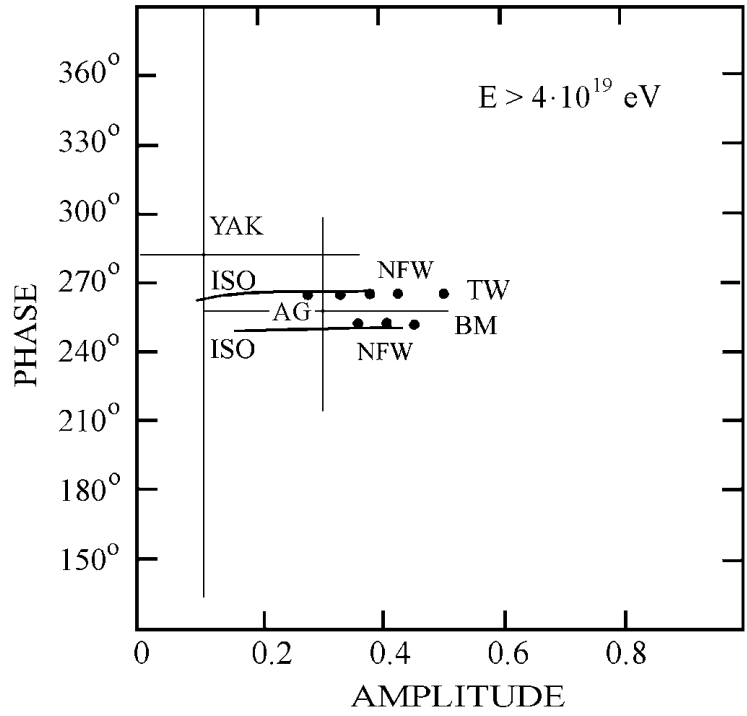
\end{document}